\documentclass[preprint]{aastex}
\usepackage{graphicx}
\usepackage{amsmath}

\shorttitle{Quick Distance Measures for Asteroids}
\shortauthors{Heinze et al.}

\begin{document}

\title{Precise Distances for Main-Belt Asteroids in Only Two Nights}

\author{Aren N. Heinze\altaffilmark{1,2} and Stanimir Metchev\altaffilmark{3,4,1}}

\altaffiltext{1}{Physics \& Astronomy Department, Stony Brook University, Stony Brook, NY 11794-3800, USA; aren.heinze@stonybrook.edu}
\altaffiltext{2}{Visiting astronomer, Kitt Peak National Observatory, National Optical Astronomy Observatory, which is operated by the Association of Universities for Research in Astronomy (AURA) under a cooperative agreement with the National Science Foundation.}
\altaffiltext{3}{Physics \& Astronomy Department, The University of Western Ontario, London, ON N6A 3K7, Canada; smetchev@uwo.ca}
\altaffiltext{4}{Centre for Planetary and Space Exploration, The University of Western Ontario, London, ON N6A 3K7, Canada; smetchev@uwo.ca}

\begin{abstract}
We present a method for calculating precise distances to asteroids using only two nights of data from a single location --- far too little for an orbit --- by exploiting the angular reflex motion of the asteroids due to Earth's axial rotation. We refer to this as the rotational reflex velocity method. While the concept is simple and well-known, it has not been previously exploited for surveys of main-belt asteroids. We offer a mathematical development, estimates of the errors of the approximation, and a demonstration using a sample of 197 asteroids observed for two nights with a small, 0.9-meter telescope. This demonstration used digital tracking to enhance detection sensitivity for faint asteroids, but our distance determination works with any detection method. Forty-eight asteroids in our sample had known orbits prior to our observations, and for these we demonstrate a mean fractional error of only 1.6\% between the distances we calculate and those given in ephemerides from the Minor Planet Center. In contrast to our two-night results, distance determination by fitting approximate orbits requires observations spanning 7--10 nights. Once an asteroid's distance is known, its absolute magnitude and size (given a statistically-estimated albedo) may immediately be calculated. Our method will therefore greatly enhance the efficiency with which 4-meter and larger telescopes can probe the size distribution of small (e.g. 100 meter) main belt asteroids. This distribution remains poorly known, yet encodes information about the collisional evolution of the asteroid belt --- and hence the history of the Solar System.

\end{abstract}


\section{Introduction}

The main asteroid belt is a relic from the formation of the Solar System. Although much of its mass has been lost, it retains a great deal of information about Solar System history and presents us with a laboratory in which we can study collisional processes that once operated throughout the circumsolar disk in which Earth and the other planets were formed. One of the most straightforward observables constraining such processes is the asteroid belt's size-frequency distribution (SFD; Bottke et al. 2005a). The current main belt's SFD can be successfully modeled as the result of 4.5 billion years of collisional evolution \citep{Bottke2005a,deElia2007}. While such models fit the `collisional wave' set up by 100 km asteroids able to survive unshattered through the age of the Solar System, they cannot be observationally tested in the 100 meter size range. 

Objects in this size range are very interesting, because they supply most near-Earth asteroids and meteorites by shattering one another and/or migrating inward via Yarkovsky and resonance effects \citep{Farinella1994,Bottke2005b}. Modern 8-10 meter telescopes can detect them, but monitoring them over many nights to determine an orbit requires a prohibitively large time investment for such powerful telescopes  (e.g., 7--10 nights; Gladman et al. 2009).  Thus their distances and sizes remain unknown, and detailed analyses are confined to larger objects \citep{Gladman2009} or use only rough statistical distances \citep{Yoshida2003,Yoshida2007}.

We present a method to obtain precise distances to main belt asteroids (MBAs) using only two nights of observations. Distances translate directly into absolute magnitudes and hence to sizes given a reasonable assumption for the albedo distribution. This method, which we refer to as rotational reflex velocity (RRV), will greatly increase the efficiency of surveys aimed at probing collisional evolution in the Solar System by measuring the SFDs for extremely small MBAs. 

We demonstrate RRV distance determination using a data set from the 0.9-meter WIYN telescope\footnote{The WIYN Observatory is a joint facility of the University of Wisconsin-Madison, Indiana University, the National Optical Astronomy Observatory and the University of Missouri.}, which we have analyzed using digital tracking \citep{digitracks} in order to enhance our sensitivity to faint asteroids.  Digital tracking is a method for detecting faint moving objects that was first applied to the Kuiper Belt (e.g. Bernstein et al. 2004), and very recently has begun to be applied to asteroids \citep{Zhai2014,digitracks}. Although the RRV distances we calculate herein are all based on our digital tracking analysis, the RRV method is equally useful for asteroids detected by more conventional means, or by other specialized methods such as those of \citet{Milani1996} and \citet{Gural2005}. 

\section{Distances from Rotational Reflex Velocity} \label{sec:math}
Suppose that at a given instant, an asteroid located a distance $d$ from an Earth-based observer is moving with velocity $v_a$, while the observer is moving with velocity $v_o$ (e.g., the orbital velocity of the Earth). The angular velocity at which the observer sees the asteroid move relative to distant stars is given by:

\begin{equation} \label{eq:bot}
\omega =  \frac{v_{a\bot} - v_{o\bot}}{d}
\end{equation}

\noindent where the $\bot$ subscript indicates the vector component perpendicular to the line of sight, so that $v_{a\bot} - v_{o\bot}$ is the projection of the asteroid's relative velocity onto the plane of the sky. 

Although $v_o$ can be accurately calculated for any Earth-based observation, the velocity $v_a$ of a newly discovered asteroid is always unknown initially, and hence the distance cannot be calculated by simply plugging the measured value of $\omega$ into Equation \ref{eq:bot}. Given appropriate measurements, however, we can isolate the component of $\omega$ that reflects the observer's motion around the geocenter due to Earth's rotation, and from this calculate the distance. This is the essence of the RRV method for distance determination.

The velocity $v_o$ of an observer on the surface of the Earth can be expressed as the sum of Earth's orbital velocity $v_{orb}$ and the velocity $v_{rot}$ with which the Earth's axial rotation carries the observer around the geocenter. Neglecting the slight asphericity of the Earth, $v_{rot} = v_{eq} \cos \theta $, where $\theta$ is the observer's terrestrial latitude and $v_{eq}$ is the Earth's equatorial rotation velocity of 1674.4 km/hr. For convenience, we define $v_{rel}$ as the asteroid's velocity relative to the geocenter: $v_{rel} = v_a - v_{orb}$. The angular velocity $\omega_g$ that would be measured by an ideal observer located at the geocenter then depends only on $v_{rel}$ and the distance, but the angular velocity $\omega_o$ that is measured by a real observer based on the Earth's surface depends also on $v_{rot}$. The two angular velocities are given by:

\begin{equation} \label{eq:vel1a}
\omega_g = \frac{v_{rel\bot}}{d}
\end{equation}

\begin{equation} \label{eq:vel1b}
\omega_o = \frac{v_{rel\bot} - v_{rot\bot}}{d}
\end{equation}

If we could measure $\omega_g$, we could therefore calculate the distance:

\begin{equation} \label{eq:vel2}
d = \frac{v_{rot}}{\omega_g - \omega_o}
\end{equation}

\noindent where we have dropped the $\bot$ subscript, because it will henceforward apply to all physical velocities in our calculations.

Now suppose that the asteroid is observed near midnight on two different nights, that the two observations are separated by exactly one sidereal day, and that the position and angular velocity $\omega_o$ are recorded for each observation. The angular distance the asteroid moved between the two observations will thus be accurately known; call this $\phi$. Because exactly one full rotation of the Earth elapsed between the two observations, the observer's position relative to the geocenter is the same for both of them.  Thus, the average geocentric angular velocity of the asteroid in between the two measurements is $\omega_g = \phi/\Delta t$, where $\Delta t$ is the elapsed time between the observations: one sidereal day.

Let the measured values of $\omega_o$ on the first and second nights be $\omega_{o1}$ and $\omega_{o2}$, and similarly let the perpendicular rotational velocities (which are obtained by calculation, not measurement) be $v_{rot1}$ and $v_{rot2}$. We can then evaluate the difference between geocentric and observer angular velocities twice: the average of $\omega_g - \omega_{o1}$ and $\omega_g - \omega_{o2}$ will be a factor of $\sqrt{2}$ more precise than a single measurement if the uncertainty on $\omega_g$ measurement is much smaller than on $\omega_{o1}$ and $\omega_{o2}$.  This is likely to be the case, since $\omega_g$ is based on a longer temporal baseline. The distance is then given by:

\begin{equation} \label{eq:av01}
d = \frac{v_{rot1} + v_{rot2}}{(\omega_g - \omega_{o1}) + (\omega_g - \omega_{o2})}
\end{equation}

So far we have assumed that $v_{rel}$ and $d$ show no appreciable change over the 24 hour period of the measurements, so that $\omega_g - \omega_{o1}$ and $\omega_g - \omega_{o2}$ are effectively two measurements of the same quantity. We will now determine the errors that result when this assumption is violated. We will first consider changes in $v_{rel}$ and $d$ that are linear in time: that is, when the first time derivatives $\dot v_{rel}$ and $\dot d$  are significant but the second derivatives $\ddot v_{rel}$ and $\ddot d$ are not.

We will parameterize the change in $v_{rel}$ by $\Delta v = (v_{rel2} - v_{rel1})/2$, and the change in $d$ by $\epsilon = (d_2-d_1)/(2d)$.  For convenience, we will also use $\delta = (v_{rot2}-v_{rot1})/(2 v_{rot})$ since the observations do not have to be taken exactly one sidereal day apart, and hence the Earth's projected rotational velocity may vary. For well-optimized observations $\delta$ should always be close to zero. Note that while $\epsilon$ and $\delta$ are unit-less, fractional changes, $\Delta v$ still has units of linear velocity. We then obtain:

\begin{equation} \label{eq:n0102}
\begin{array}{lcl}
\omega_g - \omega_{o1} &=& \frac{v_{rel}}{d} - \frac{v_{rel} - \Delta v - (1-\delta)v_{rot}}{(1-\epsilon)d}
\\ \\
\omega_g - \omega_{o2} &=& \frac{v_{rel}}{d} - \frac{v_{rel} + \Delta v - (1+\delta)v_{rot}}{(1+\epsilon)d}
\\
\end{array}
\end{equation}

We substitute these into Equation \ref{eq:av01}. Retaining terms only to second order in the variational quantities $\Delta v$, $\epsilon$, and $\delta$, after a good deal of algebra we find that the fractional error on the calculated distance is:

\begin{equation} \label{eq:err02}
\frac{d_{calc} - d_{true}}{d_{true}} = \epsilon^2 \frac{v_{rel}}{v_{rot}} - \epsilon \frac{\Delta v}{v_{rot}} + \epsilon \delta - \epsilon^2
\end{equation}

For main belt asteroids $\epsilon$ is always very small (i.e. $<10^{-3}$). This renders $\delta$ innocuous even though it can reach 0.3 for poorly-planned observations (Section \ref{sec:nonopt}). The velocity change $\Delta v$ approaches zero for asteroids at opposition, where the accuracy of Equation \ref{eq:av01} can be better than $10^{-5}$, but even months from opposition the error remains only of order $10^{-3}$. Equation \ref{eq:av01} is therefore very accurate under all conditions where the required input data could reasonably be obtained, and the uncertainties even of highly accurate measurements (Section \ref{sec:measprec}) will generally be larger than the equation's intrinsic error.

We caution the reader against a much less accurate alternative to Equation \ref{eq:av01} that may appear more attractive because it is obtained simply by applying Equation \ref{eq:vel2} individually to each night and then averaging the resulting two distance measurements:

\begin{equation} \label{eq:av00}
d = 0.5 \frac{v_{rot1}}{\omega_g - \omega_{o1}} + 0.5 \frac{v_{rot2}}{\omega_g - \omega_{o2}}
\end{equation}

\begin{figure} 
\plottwo{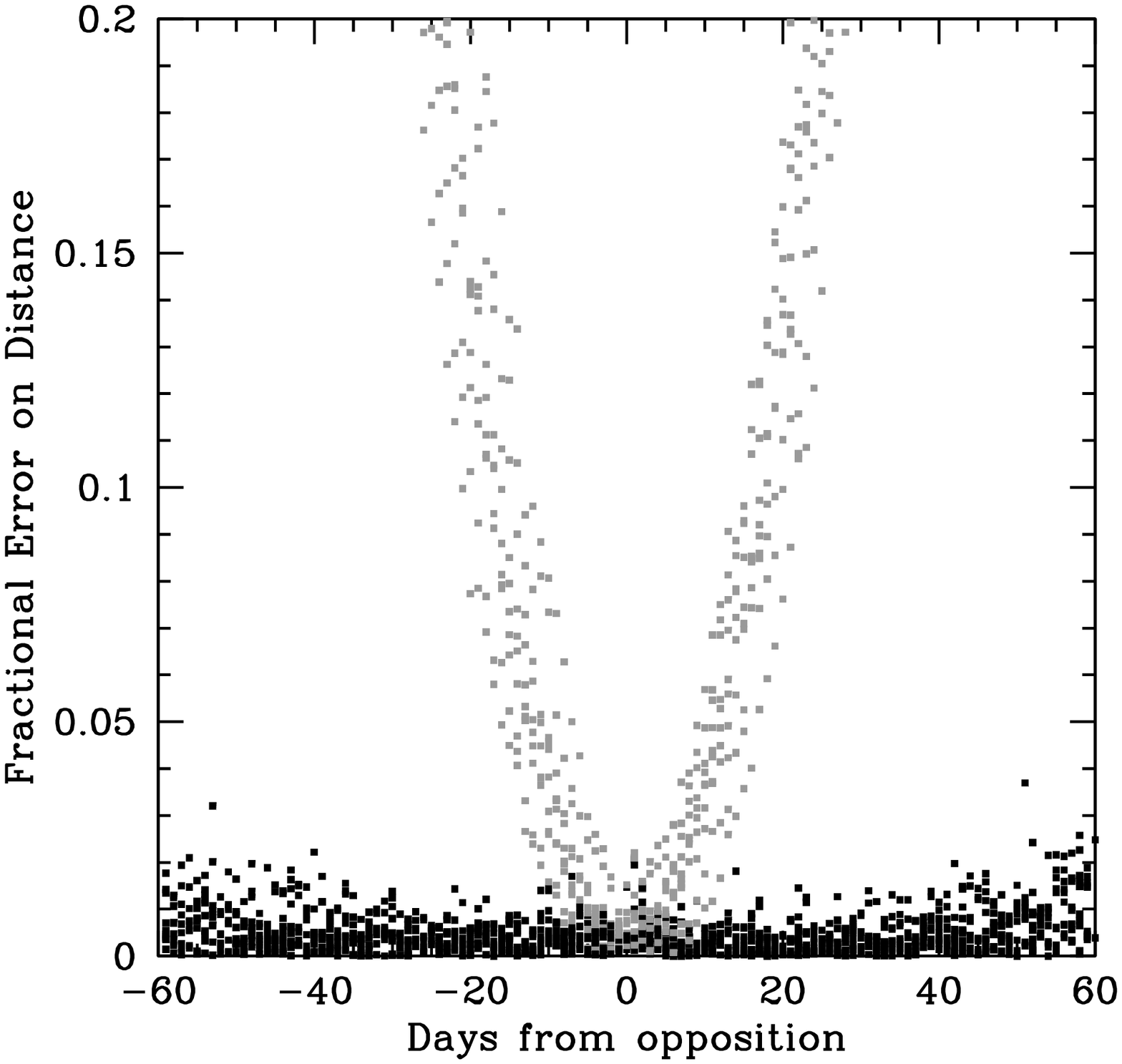}{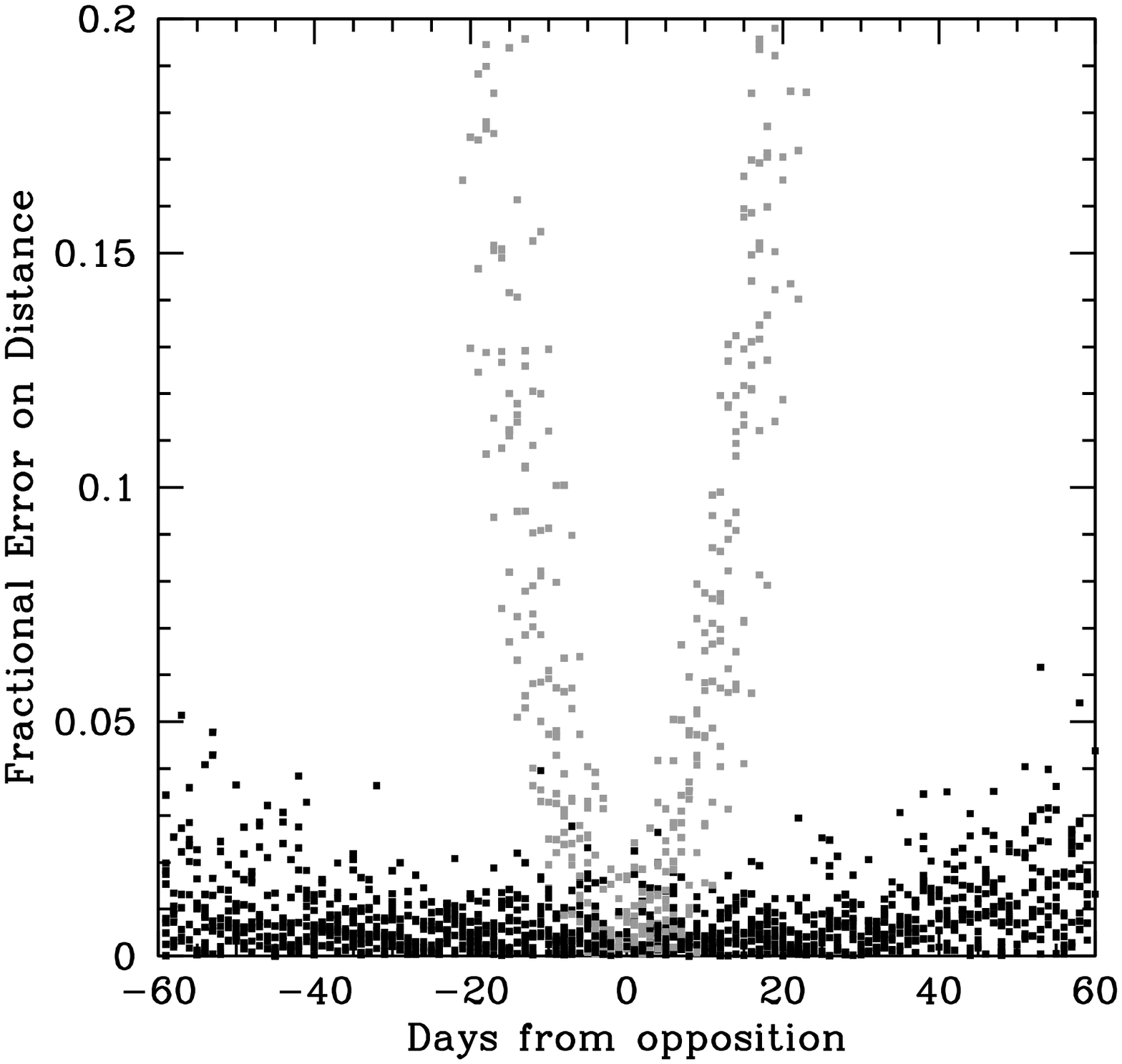}
\caption{Errors of RRV distances calculated from JPL Horizons ephemerides for representative known MBAs. Black points are for Equation \ref{eq:newav01}, a version of Equation \ref{eq:av01} that is generalized for non-optimal timing of observations (Section \ref{sec:nonopt}), while gray points are for a similarly generalized form of Equation \ref{eq:av00}. \textbf{Left:} Near optimally timed observations with $\chi$ (Equation \ref{eq:meas08}) equal to -0.0007, corresponding to $t_{2m} - t_{1m}$ equal to one minute less than a sidereal day. \textbf{Right:} Non-optimally timed observations with $\chi = 0.052$, corresponding to $t_{2m} - t_{1m}$ equal to 1.3 hours more than a sidereal day. Equation \ref{eq:newav01} yields accurate distances even with poorly timed observations, while Equation \ref{eq:av00} should never be used. `Noise' in the plots is due to roundoff error in the ephemerides rather than intrinsic error in the equations. It is more severe in the right-hand case because of the shorter temporal baseline on each night (1.7 hr vs. 2.9 hr). 
\label{fig:reflexerrors}}
%Note: Ten asteroids were used to make this figure.  They were
%asteroids number 120000 - 120009. Oppositions in either 2014 or 2015 were
%used for all objects. See March 17-18, 2015 notebook entries.
\end{figure}

\noindent Despite its intuitive simplicity, this equation should never be used. Its fractional error expansion contains the term $\Delta v^2/v_{rot}^2$, which has very large values for objects more than a few days from opposition, as illustrated by Figure \ref{fig:reflexerrors}. By contrast, for Equation \ref{eq:av01} the precursors of the $\Delta v^2/v_{rot}^2$ term and several other error terms cancel in the denominator, leaving only the very much smaller error terms given by Equation \ref{eq:err02}. Equation \ref{eq:av00} should therefore be avoided, and Equation \ref{eq:av01} (or its generalized form, Equation \ref{eq:newav01}) should always be used for RRV distances.

%For realistic MBA orbits, the contribution of the second derivative $\ddot v_{rel}$ is only at the $10^{-3}$ level, and that from $\ddot d$ is many times smaller. Since distances accurate to within 2\% are completely sufficient for studies of MBA size statistics, the error of our approximation is negligible in this context.

\section{Asteroid Angular Velocities from Positions Measured at Discrete Times} \label{sec:discrete}

\subsection{Basic Equations and Methodology} \label{sec:opt}
For simplicity, the discussion in Section \ref{sec:math} assumed the observer could make instantaneous angular velocity measurements. In practice, of course, angular velocity measurements must be made over a period of time.  This has very little effect on our mathematical derivation. The only change is that the projected velocities of the observer relative to the geocenter, $v_{rot1}$ and $v_{rot2}$, should not be considered as instantaneous but rather as averages over the same period of time in which the angular velocities $\omega_{o1}$ and $\omega_{o2}$ were measured.  

Positions and angular velocities of asteroids are measured in terms of right ascension (RA) and declination (DEC) on the celestial sphere, and thus are two dimensional.  For example, an observer could measure an asteroid as having an angular velocity of -37.36 arcsec/hr in RA (the negative sign indicating westward rather than eastward motion) and 10.32 arcsec/hr in DEC, with the positive sign indicating the object is moving toward celestial north. In principle, independent versions of Equation \ref{eq:av01} could be constructed for each dimension, with angular velocities in RA corresponding exactly to projected east-west motion of the observer and DEC velocities corresponding to projected north-south motion. However, the true rotational velocity of the observer is always strictly east-west, and although its projection $v_{rot}$ can have a north-south component, this component is usually small --- i.e., less than 10\% of the east-west component for observations with mean hour angle in the range $\pm1.0$ targeting asteroids within $20^{\circ}$ of the celestial equator. As we will find below, for our test observations only the RA velocity components are large enough to supply useful distances. We expect this will generally be the case.

In most asteroid surveys, many fields are observed, and each field is visited at least three times per night.  Each visit yields a single celestial position of each detected asteroid. Two observations of each asteroid on each night will suffice for our purposes.  Using the traditional symbols $\alpha$ and $\delta$ for RA and DEC, suppose that on night 1 we obtain observations $\alpha_{1a},\delta_{1a}$ at time $t_{1a}$ and $\alpha_{1b},\delta_{1b}$ at time $t_{1b}$, and similarly on night 2. Then we can also define mean positions and the mean time $\alpha_{1m} = (\alpha_{1a}+\alpha_{1b})/2$, $\delta_{1m} = (\delta_{1a}+\delta_{1b})/2$, and $t_{1m} = (t_{1a}+t_{1b})/2$, with the analogous quantities being calculated for night 2. Note that the times here are measured on a continuous sequence and not reset when the date changes. For simplicity, at present we will suppose further that the time of each measurement on night 2 is exactly one sidereal day later than the time of the corresponding measurement on night 1: thus $t_{2a} - t_{1a} = t_{2b} - t_{1b} = $ one sidereal day. The required angular velocities are then given by the following equations, where for simplicity we show only the RA component:

\begin{equation} \label{eq:meas01}
\omega_{o1\alpha} = \frac{\alpha_{1b} - \alpha_{1a}}{t_{1b} - t_{1a}}
\end{equation}

\begin{equation} \label{eq:meas02}
\omega_{o2\alpha} = \frac{\alpha_{2b} - \alpha_{2a}}{t_{2b} - t_{2a}}
\end{equation}

\begin{equation} \label{eq:meas03}
\omega_{g\alpha} = \frac{\alpha_{2m} - \alpha_{1m}}{t_{2m} - t_{1m}}
\end{equation}

\noindent where the denominator of the last equation is of course also equal to one sidereal day.

\begin{figure} 
\includegraphics[scale=0.8]{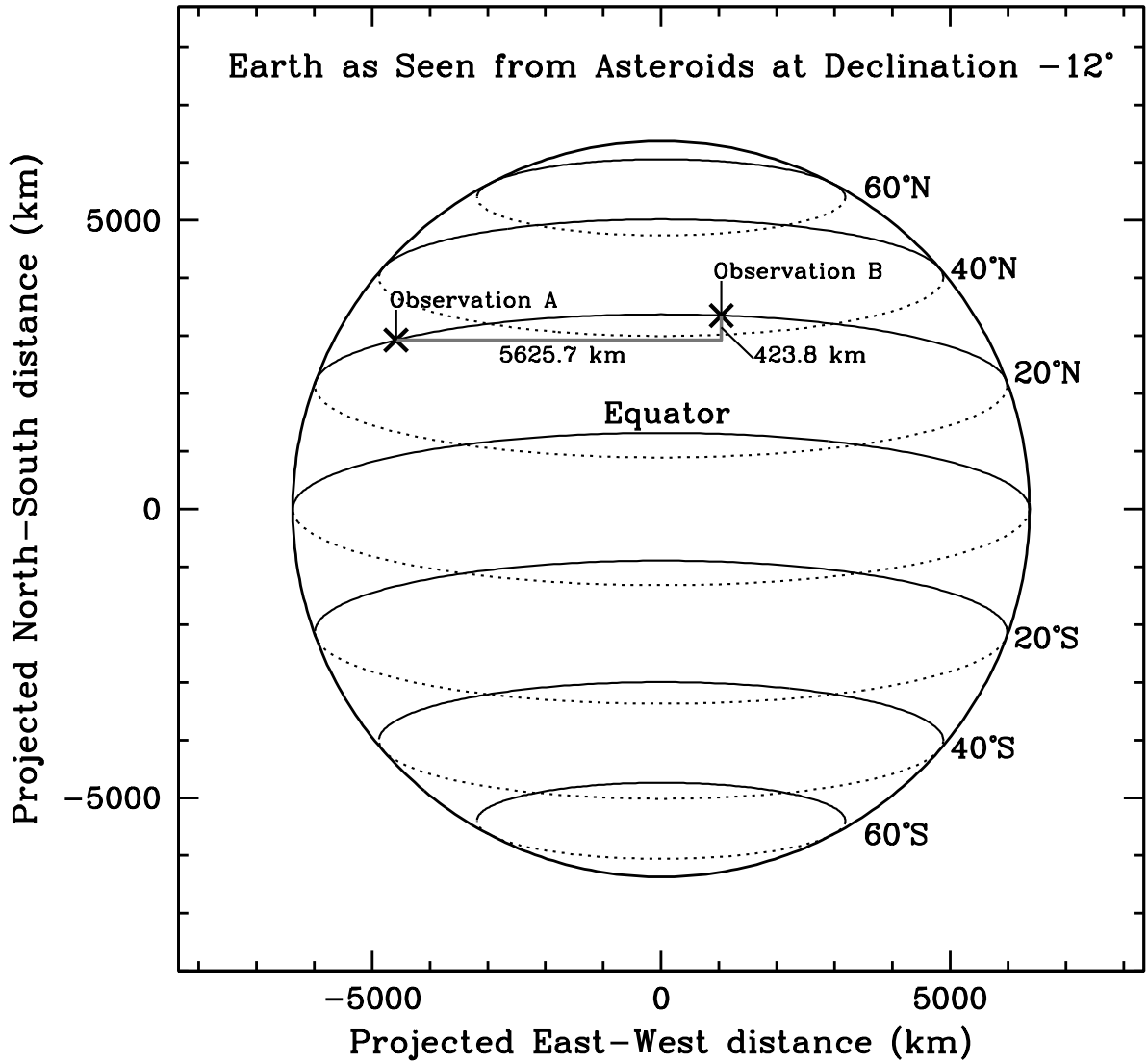}
\caption{Example illustration of an observer's motion due to the Earth's rotation, as viewed from an asteroid.  Observation A and Observation B are both made from the same location on Earth's surface (at 20$^{\circ}$ N, roughly the latitude of Mauna Kea), but they are separated by an elapsed time of 3.989 hr, corresponding to 60$^{\circ}$ of sidereal rotation. The rotational velocity of Earth's surface at 20$^{\circ}$ N lat is 1573.4 km/hr. The effect on the asteroid's measured positions is determined by the projected distance moved between the observations, which has eastward and northward components as labeled. The projected rotational velocity $v_{rot}$ that should be used in the RRV distance calculation (Equation \ref{eq:av01} or Equation \ref{eq:newav01}) is therefore $5625.7 \hspace{0.05in} \mathrm{km}/3.989 \hspace{0.05in} \mathrm{hr} = 1410.3$ km/hr eastward and $423.8 \hspace{0.05in} \mathrm{km}/3.989 \hspace{0.05in} \mathrm{hr} = 106.2$ km/hr northward.
\label{fig:earth}}
\end{figure}
%1674.4
%220 -4586.141	2928.290
%280 1039.594	3352.077
%86164.09054

The only other quantities needed for the distance calculation are $v_{rot1}$ and $v_{rot2}$, the observer's projected rotational velocities relative to the geocenter. As illustrated by Figure \ref{fig:earth}, the observer's projected motion relative to the geocenter traces out a segment of an ellipse whose axis ratio is the sine of the asteroid's declination. The projected distance of the observer from the geocenter is given by:

\begin{equation} \label{eq:proj1}
\begin{array}{lcl}
x (t) &=& R_{\earth}\cos (\theta_{lat}) \sin \left(\frac{2 \pi (t - t_z)}{\tau_S} \right) \\
\\
 y (t) &=& -R_{\earth}\cos (\theta_{lat}) \sin(\delta) \cos \left(\frac{2 \pi (t - t_z)}{\tau_S} \right) + R_{\earth}\sin (\theta_{lat})
\end{array}
\end{equation}

\noindent where $x$ and $y$ are positive toward the east and north; $\theta_{lat}$ is the observer's latitude on the Earth; $\delta$ is the declination of the asteroid; $t$ is the time of the observation; $t_z$ is the time when the asteroid stands at zero hour angle for the observer; and $\tau_S$ is the length of a sidereal day. Note that the argument $2 \pi (t - t_z)/\tau_S$ is the hour angle at which the observer sees the asteroid: at $t=t_z$ the asteroid crosses the meridian and stands closest to the observer's zenith and in the best sky position for observing. 

For observations at $t_{1a}$ and $t_{1b}$, the projected linear velocity $v_{rot1}$ in the east-west ($x$) and north-south ($y$) directions is:

\begin{equation} \label{eq:proj2x}
v_{rot1x} = \frac{x (t_{1b}) - x (t_{1a})}{t_{1b} - t_{1a}}
\end{equation}

\begin{equation} \label{eq:proj2y}
v_{rot1y} = \frac{y (t_{1b}) - y (t_{1a})}{t_{1b} - t_{1a}}
\end{equation}

\noindent Figure \ref{fig:earth} illustrates a specific example of two observations of an asteroid at -12$^{\circ}$ DEC, made from a single observing site at 20$^{\circ}$ N lat on the Earth's surface.  The first observation is made when the asteroid is rising, and the second about four hours later, shortly after it has crossed the meridian. The rotational velocity of Earth at 20$^{\circ}$ N lat is 1573.4 km/hr and is of course strictly eastward, but projecting the velocity into a plane perpendicular to the line of sight reduces the absolute magnitude and introduces a north-south component (except for asteroids exactly at 0$^{\circ}$ DEC). In this example the projected velocities east and north are 1410.3 km/hr and 106.2 km/hr.   Since celestial RA and DEC correspond exactly to terrestrial longitude and latitude, the east-west physical velocity of Equation \ref{eq:proj2x} can be combined with the RA components of the angular velocities from Equations \ref{eq:meas01}--\ref{eq:meas03} and plugged directly into Equation \ref{eq:av01} to yield the distance. 

We will now provide a simplified but basically realistic example. Suppose the MBA illustrated by Figure \ref{fig:earth} is being observed at opposition.  It will be in its retrograde loop, and hence its motion in RA will be negative (westward). The angular velocity $\omega_{o}$ measured by the observer will be faster to the west than the geocentric angular velocity $\omega_g$, because on the night side of the Earth the rotational velocity adds to the orbital velocity, making the asteroid fall back westward at a faster rate. Suppose $\omega_{g\alpha} = -35$ arcsec/hr and $\omega_{o1\alpha} = -36$ arcsec/hr. The rotational reflex velocity, given by the difference $\omega_{g\alpha} - \omega_{o1\alpha}$, is then 1 arcsec/hr or $4.85 \times 10^{-6}$ rad/hr. Taking the physical and angular velocities to be exactly the same on the second night for simplicity, Equation \ref{eq:av01} reduces to:

\begin{equation} \label{eq:calc01}
d = \frac{v_{rotx}}{\omega_{g\alpha} - \omega_{o\alpha}} = \frac{1410.3 \hspace{0.05in} \mathrm{km/hr}}{4.85 \times 10^{-6} \hspace{0.05in} \mathrm{rad/hr}} = 2.91 \times 10^8 \hspace{0.05in} \mathrm{km} = 1.94 \hspace{0.05in} \mathrm{AU}
\end{equation}

With Equation \ref{eq:calc01} we have calculated the distance to an asteroid based on the known properties of Earth's rotation. Interestingly, \citet{Bernstein2000} have developed a closely analogous method for determining the distance to a Kuiper Belt object using the known properties of Earth's \textit{orbit}. Just as our derivation does not (explicitly) include the effects of Earth's orbit, that of \citet{Bernstein2000} does not include the effects of Earth's rotation. Neglecting Earth's orbit works in our case because, as we have demonstrated in Section \ref{sec:math}, the dominant errors due to orbital acceleration of both the Earth and the asteroid cancel when the correct form of the distance equation is used. Similarly, \citet{Bernstein2000} can safely neglect Earth's rotation because the Kuiper Belt objects whose distances they aim to measure are sufficiently far away that their RRV signature is negligible. It follows that the method of \citet{Bernstein2000} should not be directly applied to measuring asteroid distances without some form of correction for the RRV signal, although (as they point out) it is useful for distinguishing true Kuiper Belt objects from asteroids that have similar angular velocities for a few days near their turnaround points.

\subsection{The Case of Observations that are Not Optimally Timed} \label{sec:nonopt}
Up to now we have assumed for simplicity that the observations obtained on successive nights are exactly one sidereal day apart.  While it is not difficult to plan and execute observations that satisfy this criterion to within a few minutes, clouds or instrument failures could intervene, or one could be processing an archival data set that was not taken with RRV distances in mind. In such cases the mean times $t_{1m}$ and $t_{2m}$ may be separated by up to a couple of hours less or more than one sidereal day. A direct calculation of the geocentric angular velocity $\omega_g$ then becomes impossible.  In its place, we must use instead the mean angular velocity across the two nights, defined by:

\begin{equation} \label{eq:meas08b}
\omega_{\mu \alpha} = \frac{\alpha_{2m} - \alpha_{1m}}{t_{2m} - t_{1m}}
\end{equation}

We will define three additional quantities.  The time $t_s$ is the moment exactly one sidereal day before $t_{2m}$ (in the ideal case, $t_{1m}$ would have been equal to $t_s$). The projected rotational velocity $v_{rot,s}$ is the observer's mean projected rotational velocity over the time interval between $t_{1m}$ and $t_s$.  Finally, the fractional parameter $\chi$ is given by:

\begin{equation} \label{eq:meas08}
\chi = \frac{t_s - t_{1m}}{t_{2m} - t_{1m}}
\end{equation}

\noindent Note that $\chi$ goes to zero in the case of optimally-timed measurements.

The distance formula analogous to Equation \ref{eq:av01} becomes:

\begin{equation} \label{eq:newav01}
d = \frac{v_{rot1} + v_{rot2} - 2\chi v_{rot,s}}{(\omega_{\mu}  - \omega_{o1}) + (\omega_{\mu}  - \omega_{o2})}
\end{equation}

The error terms for Equation \ref{eq:newav01} are the same as those given in Equation \ref{eq:err02}, with the addition of an $\epsilon^2 \chi$ term. For main belt asteroids, this will always be negligible. In closing our discussion of error in the case of non-optimal observation timing, we note that $\delta$, the fractional change in projected rotational velocity from night 1 to night 2, is likely to be considerably larger in this case, up to 0.3.  However, in the error terms for Equation \ref{eq:newav01} it is always multiplied by $\epsilon$, which is always of order $10^{-3}$ or smaller for MBAs, and thus should not constitute a major source of error. Equation \ref{eq:newav01} is therefore the central formula for asteroid distance determination with RRV.

In addition to our analytical error calculation (Equation \ref{eq:err02}), we have probed the errors of our approximations numerically using ephemerides of known objects from the JPL Horizons ephemeris generator. Note that no actual observations are involved in this test, only ephemeris positions and velocities for objects with well-known orbits. In Figure \ref{fig:reflexerrors} we plot the error of Equation \ref{eq:newav01} as determined by this test. In contrast to our analytical error calculations, this test intrinsically includes the contributions of second and higher-order time derivatives of $d$ and $v_{rel}$, since it uses JPL Horizons ephemerides for real objects. The plotted errors are affected by roundoff error in the ephemerides (accurate only to 0.15 arcsec in RA), which introduces pseudo-noise such that only the lower envelope of the plotted points can be meaningfully compared with our analytical error estimates. Nevertheless, the plots show that Equation \ref{eq:newav01} produces accurate distances out to at least 60 days from opposition; that accurate distances can be obtained from observations that are not optimally timed; and that the second-order time derivatives $\ddot v_{rel}$ and $\ddot d$ do not introduce significant error.

\subsection{Measurement Precision Required for Good Distances} \label{sec:measprec}
Figure \ref{fig:reflexerrors}, although based on calculated ephemerides and not actual data, shows that relatively small position errors can create distance inaccuracies at the level of a few percent.  We now consider the measurement precision that is necessary to yield distances to various levels of accuracy.

In Equation \ref{eq:calc01}, the difference between the night-to-night mean angular velocity in RA ($\omega_{\mu\alpha}$) and the angular velocity in RA observed within a given night ($\omega_{o\alpha}$) was one arcsec/hr, and this rotational reflex velocity corresponded to a distance of about 2 AU. This is typical for observations obtained near opposition at terrestrial latitudes of 20--30$^{\circ}$. The fractional uncertainty of the distance is equal to that of $\omega_{\mu\alpha} - \omega_{o\alpha}$. This will generally be dominated by the uncertainty in $\omega_{o\alpha}$, since it is calculated within a single night and is thus based on a smaller temporal baseline than $\omega_{\mu\alpha}$. Thus, if $\omega_{o\alpha}$ is measured with an accuracy of 0.1 arcsec/hr, we should expect the distance to have an accuracy of only 10\% at 2 AU.  At 1 AU, the RRV signature is twice as large: $\omega_{\mu\alpha} - \omega_{o\alpha}$ will therefore be about 2 arcsec/hr, and the calculation will be accurate to 5\%. 

As we will now illustrate, one can usually obtain angular velocity measurements with better than 0.1 arcsec/hr uncertainties in practice. In our current paradigm, $\omega_{o\alpha}$ is based on two measurements of celestial positions separated by a time $t_b-t_a$ (i.e., Equation \ref{eq:meas01} or \ref{eq:meas02}). If the uncertainty on RA for these measurements is $\sigma_{\alpha}$, the uncertainty on $\omega_{o\alpha}$ will be:

\begin{equation} \label{eq:meas09}
\sigma_{\omega} = \frac{\sigma_{\alpha}\sqrt{2}}{t_{b} - t_{a}}
\end{equation}

Based on our experience, well-sampled (e.g. $<0.5$ arcsec/pixel) images taken in $\sim1.5$ arcsecond seeing can yield $\sigma_{\alpha}$ = 0.03--0.05 arcsec for bright objects (e.g., those at least several times brighter than the 10$\sigma$ detection threshold). For faint objects near the detection limit, uncertainties of $\sigma_{\alpha} \sim 0.1$ arcsec are more typical\footnote{Note that the distance uncertainty must be evaluated differently for digital tracking observations, where the angular velocity is directly measured (see Section \ref{sec:angdigi}) rather than being obtained from two positions.}  A temporal baseline of 5 hr is easy to obtain near opposition.  Thus, we can expect $\sigma_{\omega}$ = 0.008--0.014 arcsec/hr for bright objects and $\sigma_{\omega}$ = 0.03 arcsec/hr for fainter objects. At 2 AU, distances will therefore be accurate to about 3\% for faint objects and 1--1.5\% for bright objects, while at a distance of 1 AU, the accuracy should be about 1.5\% even for faint objects. These accuracies are sufficient for good analyses of the size statistics of small MBAs, especially given that their albedos are unknown and must be treated as a statistical distribution spanning a factor of $\sim$10 \citep{Pravec2012,Masiero2013}. Where the objects are faint or distant (e.g. 3--5 AU), or where extremely accurate distances are desired, it may be possible to obtain smaller values of $\sigma_{\alpha}$ by obtaining each RA measurement not from a single image but from the average over a set of images acquired close together in time.  Extending the same principle further, we can use the technique of digital tracking to obtain accurate measurements for objects too faint even to be detected in individual images, as we describe below.

\section{Measurements Using Digital Tracking Data} \label{sec:digitrack}

\subsection{A Brief Introduction to Digital Tracking}
The observations we use to demonstrate the usefulness of rotational reflex distances for asteroids were obtained using digital tracking, as described in our companion paper \citet{digitracks}. These data consist of 126 two-minute exposures of a single starfield that we acquired using the WIYN 0.9-meter telescope at Kitt Peak on the night of April 19, and 130 identically acquired images from the following night. 

Very briefly, a digital tracking search involves shifting and stacking such sets of images to reveal moving objects too faint to be detected on any individual frame. A separate trial stack is produced for each angular velocity vector in a finely-sampled grid that spans the full range of possible sky velocities for the target population. For MBAs, a digital tracking data set can span only one night \citep{digitracks}, hence we analyze our April 19 and April 20 observations independently. Digital tracking has been used to great effect for Kuiper Belt objects (e.g. Allen et al. 2001, Fraser \& Kavelaars 2009, and many others), but has not typically been used for faster-moving asteroids where a larger numbers of trial vectors must be probed. It is now computationally tractable even for asteroids, however, and produces a factor of $\sim 10$ increase in sensitivity over conventional methods. This sensitivity increase enables us to detect 215 asteroids within the 1-degree field of our test observations, despite using only a small, 0.9-meter telescope. We obtained precise angular velocity measurements on both nights for 197 of these asteroids, including 48 previously known objects with accurate orbits. These last allow us to test the accuracy of RRV distance measurements using real data.

\subsection{Angular Velocity Measurement with Digital Tracking} \label{sec:angdigi}
The geocentric angular velocity $\omega_g$ of an asteroid detected on two subsequent nights is determined in exactly the same way with digital tracking observations as with conventional data. The determination of the observed angular velocities $\omega_o$ on each individual night is different, however --- particularly because digital tracking enables the accurate measurement of asteroids that cannot even be detected on an individual image.

Any asteroid detection in a digital tracking search occurs on a particular trial image stack, which corresponds to a particular angular velocity.  Thus, for example, our automated digital tracking routine might search for objects with angular velocities between -50 and -20 arcsec/hr in RA and -10 and +20 arcsec/hr in DEC, using a grid spacing of 0.2 arcsec/hr, and might detect a particular asteroid as a bright point source on a trial stack whose shifts correspond to an angular velocity of -41.2 arcsec/hr in RA and 11.8 arcsec/hr in DEC.

We improve on the relatively crude angular velocity measurement that is implicit in such a detection by probing a new, much more finely spaced grid of angular velocity vectors, using small postage-stamp images centered on the detected asteroid to make the search computationally tractable.  For each trial stack in this finely sampled grid, we calculate the measured flux of the asteroid within a small aperture of radius roughly equal to the half-width of the point spread function (PSF) of our image.  We then perform a 2-D quadratic fit to the measured flux as a function of the trial angular velocity. The angular velocity at which the quadratic fit reaches its peak value then constitutes an accurate measurement of $\omega_o$ for that asteroid, and the uncertainty on $\omega_o$ is derived from the uncertainty of the quadratic fit, which is naturally larger for faint objects that are more affected by sky background noise.

\begin{figure} 
\plottwo{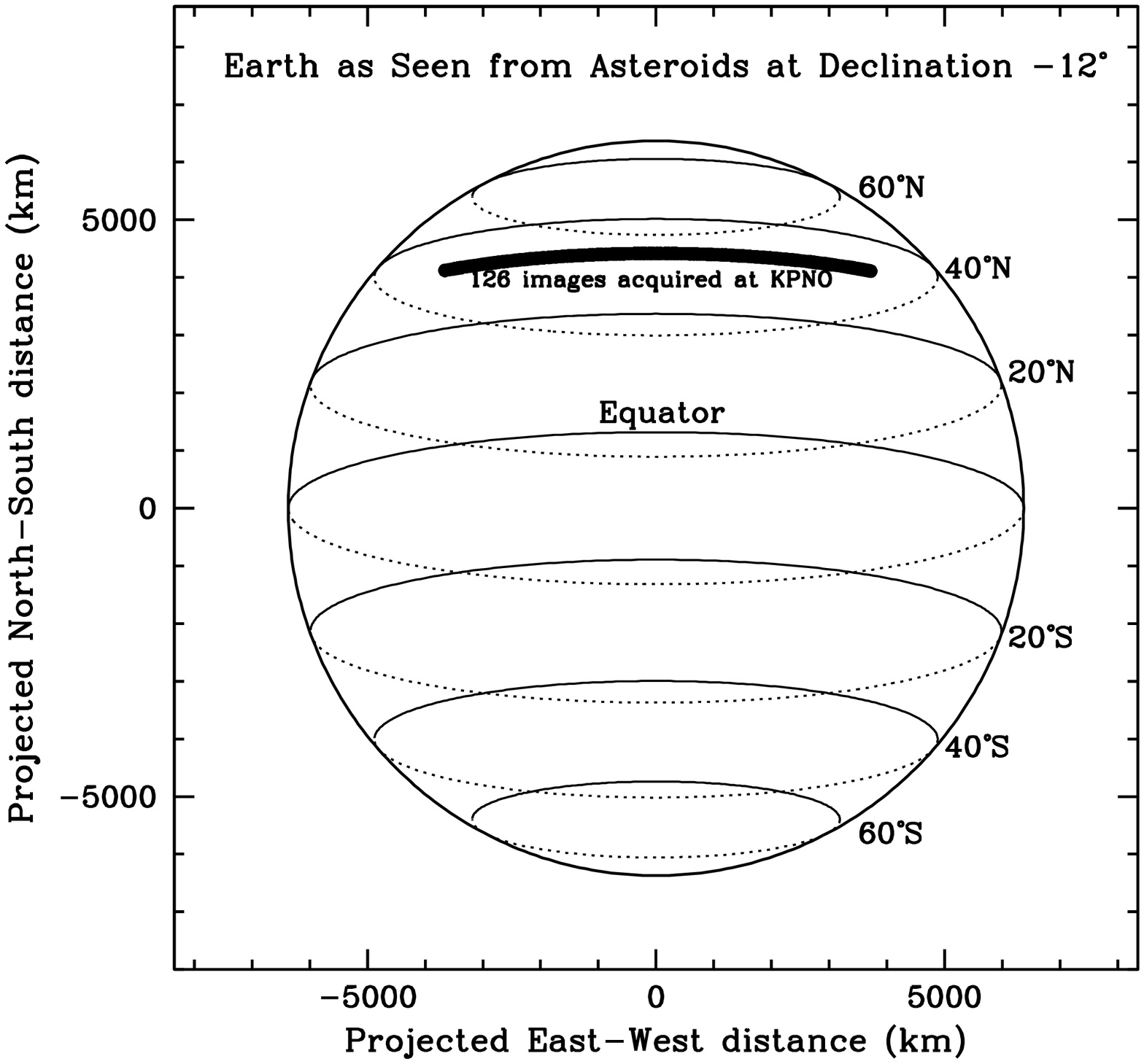}{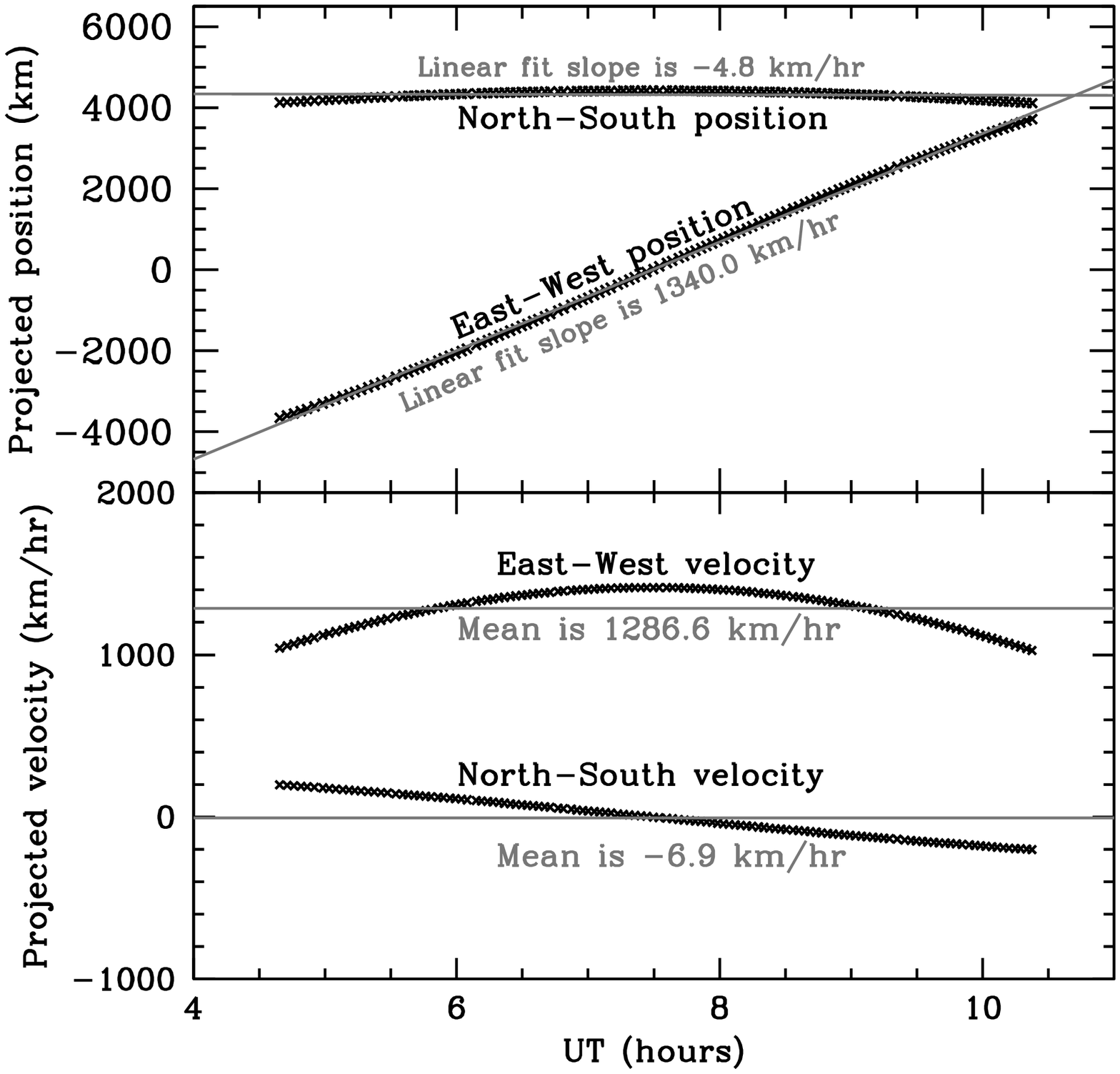}
\caption{The projected motion of a ground-based observer due to the Earth's rotation, which
affects the measured positions and angular velocities of detected asteroids.
\textbf{Left:} Schematic diagram of Earth viewed from the perspective of asteroids in our April 19, 2013
field, which was near -12$^{\circ}$ declination. The heavy black curve is composed of 126 X's
marking the projected position ($x,y$ from Equation \ref{eq:proj1}) of the Kitt Peak (31.9583$^{\circ}$ N lat) observer at the
moment each of our images was taken. \textbf{Top Right:} $x(t)$ and $y(t)$ for our April 19 images, showing the linear fit velocities that should
be used for comparison to linear, constant-velocity digital tracking results. 
\textbf{Bottom Right:} The velocities $\dot x(t)$ and $\dot y(t)$ for our data,
showing how the mean velocities are significantly different from the best-fit velocities
shown in the top panel.
\label{fig:proj1}}
\end{figure}

All the inputs required to calculate the distance using Equation \ref{eq:newav01} are now available except $v_{rot}$. As in the case of discrete observations, Equation \ref{eq:proj1} indicates how $v_{rot}$ should be calculated, but there is one subtle aspect. The appropriate value of $v_{rot}$ is not a projected distance divided by an elapsed time as it was in the discrete case (Figure \ref{fig:earth}), but neither is it the time average of the projected velocities $\dot x (t)$ and $\dot y (t)$. Instead, the value of $v_{rot}$ that corresponds to the angular velocity measured by digital tracking is given by the slopes of the best linear fits to $x$ and $y$ as functions of time. Figure \ref{fig:proj1} illustrates the difference between linear fit velocities and average velocities. 

For our data, the linear fit velocities of the Kitt Peak observer were 1340.0 km/hr eastward and -4.8 km/hr northward on April 19; and 1337.3 km/hr eastward and -16.3 km/hr northward on April 20; by contrast the average velocities were 1286.6 km/hr eastward and -6.9 km/hr northward on April 19 and 1283.0 km/hr eastward and -14.6 km/hr northward on April 20.  In this case, the north-south velocities are negligible for practical purposes.

%\begin{figure} 
%\includegraphics[scale=0.8]{driftplot04.eps}
%\caption{The measured angular velocities of asteroids detected in our April 19 data are shown as black hexagons.
%The grey squares plotted to suggest the locus of main belt asteroids are known objects over a much
%larger region of the sky (a disk 6 degrees in diameter) centered on our field at the time of our
%observations. These motions were obtained from the Minor Planet Center. The square outline
%indicates the boundary of the region of motion space that we have searched.
%Note that we have no detections in the corner regions far from the locus of known asteroids. This
%provides further evidence that none of our objects are false positives, as we have already demonstrated
%through the tests described in Section \ref{sec:mancheck}.
%\label{fig:drift}}
%\end{figure}

\section{Results from our Test Data}

\subsection{Initial Distance Determinations} \label{sec:initdist}
Our observations on 2013 April 19 span a temporal range of 5.72 hr. Those obtained on April 20 have the same range, and are centered 23.968 hr later in time.  As this interval is only 2 minutes longer than a sidereal day, the timing of our observations is almost exactly optimal as defined in Section \ref{sec:opt}: the parameter $\chi$ from Equation \ref{eq:meas08} is only 0.0014.
We detected a total of 215 asteroids through digital tracking analysis of our observations in a field of view only slightly larger than 1 square degree. We reported positions and brightnesses for all of these objects to the Minor Planet Center. Of these 215 MBAs, 202 were detected on both April 19 and April 20, and 197 of these two-night objects had sufficiently accurate measurements on both nights for good distance calculations. Of these, 48 were previously discovered objects with accurately known orbits, nine were previously discovered objects with poorly known orbits prior to our observations, and the remaining 140 were new discoveries for which we received designations from the Minor Planet Center. \citet{digitracks} provides further details.

We have used Equation \ref{eq:newav01} to calculate distances for all 197 objects with sufficient measurements.  The 48 objects in this sample with well-known orbits allow us to test the accuracy of rotational reflex distances using real data. The mean absolute error of our calculated distances for these objects is 0.042 AU, and the mean absolute fractional error is 2.2\%.  As expected, errors are smaller for more nearby objects. Distances with 2\% accuracy are more than sufficient to analyze the size statistics of MBAs.  Nevertheless, even better results are possible, as we will see below.

\subsection{Correcting for Bias from Track Curvature} \label{sec:two_precise}

Close investigation reveals some evidence of systematic error in the distance calculations described in Section \ref{sec:initdist}. The weighted average signed fractional error for the 48 known objects is $-0.0179 \pm 0.0017$: i.e., the calculated distances are 1.8\% too small on average, and the offset is 10$\sigma$ significant. 

It might be suspected that this bias is due to using the wrong value for $v_{rot}$ --- i.e., that we were wrong to use linear fit velocities rather than time-averaged velocities (Section \ref{sec:angdigi}). This is easily demonstrated not to be the case, however. Substituting averaged values in place of linear fit values for $v_{rot}$ changes the result too much and in the wrong direction: the mean systematic offset goes from -1.8\% to -5.2\%. The true cause of the bias is more subtle, and points to a further interesting measurement that can be extracted from our data.

With digital tracking, our measurement of the single-night observed angular velocity $\omega_o$ is made over a period of time equal to the temporal span of our observations on each night. The instantaneous angular velocity over this interval is continuously changing, with the rotational reflex velocity itself being the dominant source of change. For example, the rotation of the Earth causes the westward sky motions of the asteroids to be slightly faster near midnight than they are at the beginning and end of each night's observations (see Figure \ref{fig:proj1}). As we stated in Section \ref{sec:angdigi} and illustrated by Figure \ref{fig:proj1}, we can account for the acceleration and curvature of an asteroid's observed motion by using a value for $v_{rot}$ that is obtained by a linear fit to the observer's projected coordinates $x(t)$ and $y(t)$. 

This assumes, however, that the angular velocity measurement is effectively an \textit{unweighted} fit to angular position as a function of time. Such is not necessarily the case. For example, the images taken near midnight, when the westward sky motion is at its fastest, may be more sensitive due to better seeing or lower atmospheric extinction. The digital tracking stack will then effectively weight them more highly and produce a mean angular velocity that is systematically too fast.  This effect would produce a systematic underestimation of the asteroids' geocentric distances: exactly what we observe for known asteroids in our data.  

The effective weighting of different images will not necessarily be the same for different asteroids. An asteroid that passes too close to a bright star near midnight will only be measured near the beginning and end of the night, resulting in a measured angular velocity slower than the true average value.  Asteroids can also exhibit significant brightness changes during the night due to their own rotation, which will change the effective weighting of different images in the final measurement of their angular velocities. 

If asteroids actually moved in straight lines at constant velocity, none of these weighting effects would introduce error into our angular velocity determinations.  If we could \textit{predict} the curvature and acceleration of each asteroid's track due to the Earth's rotation, and apply small shifts to each image in our digital tracking stacks in order to linearize the motion, we would therefore remove the bias in our angular velocity determinations. We now describe how to do this.

The angular distances between the position an observer measures for an asteroid and the position it would have if measured from the geocenter are simply a reflection of Equation \ref{eq:proj1}, scaled by the distance:

\begin{equation} \label{eq:curve1}
\begin{array}{lcl}
\Delta \alpha (t) &=& -\displaystyle\frac{x (t)}{d} \\
\\
\Delta \delta (t) &=& -\displaystyle\frac{y (t)}{d} 
\end{array}
\end{equation}

\begin{figure} 
\plottwo{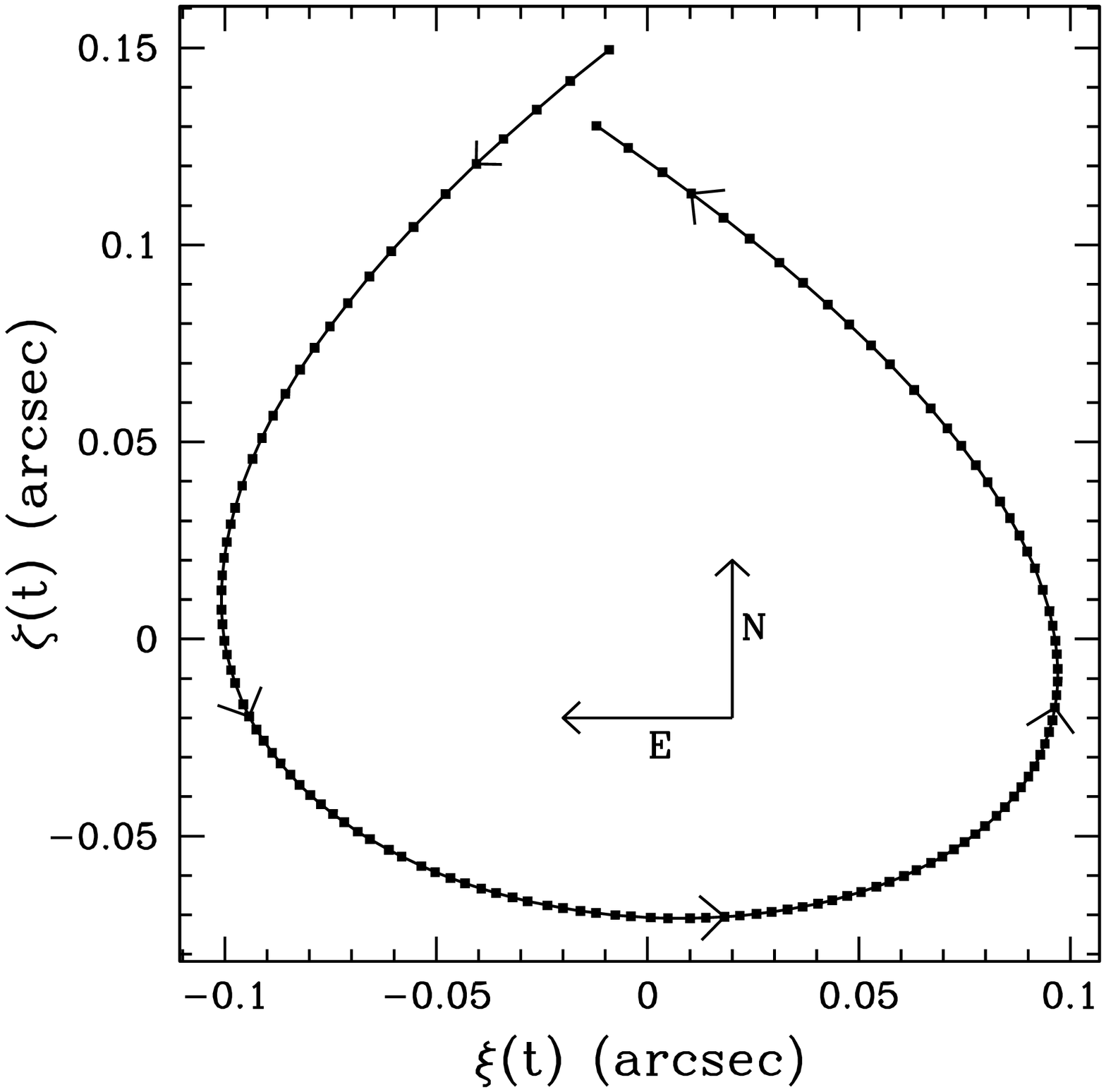}{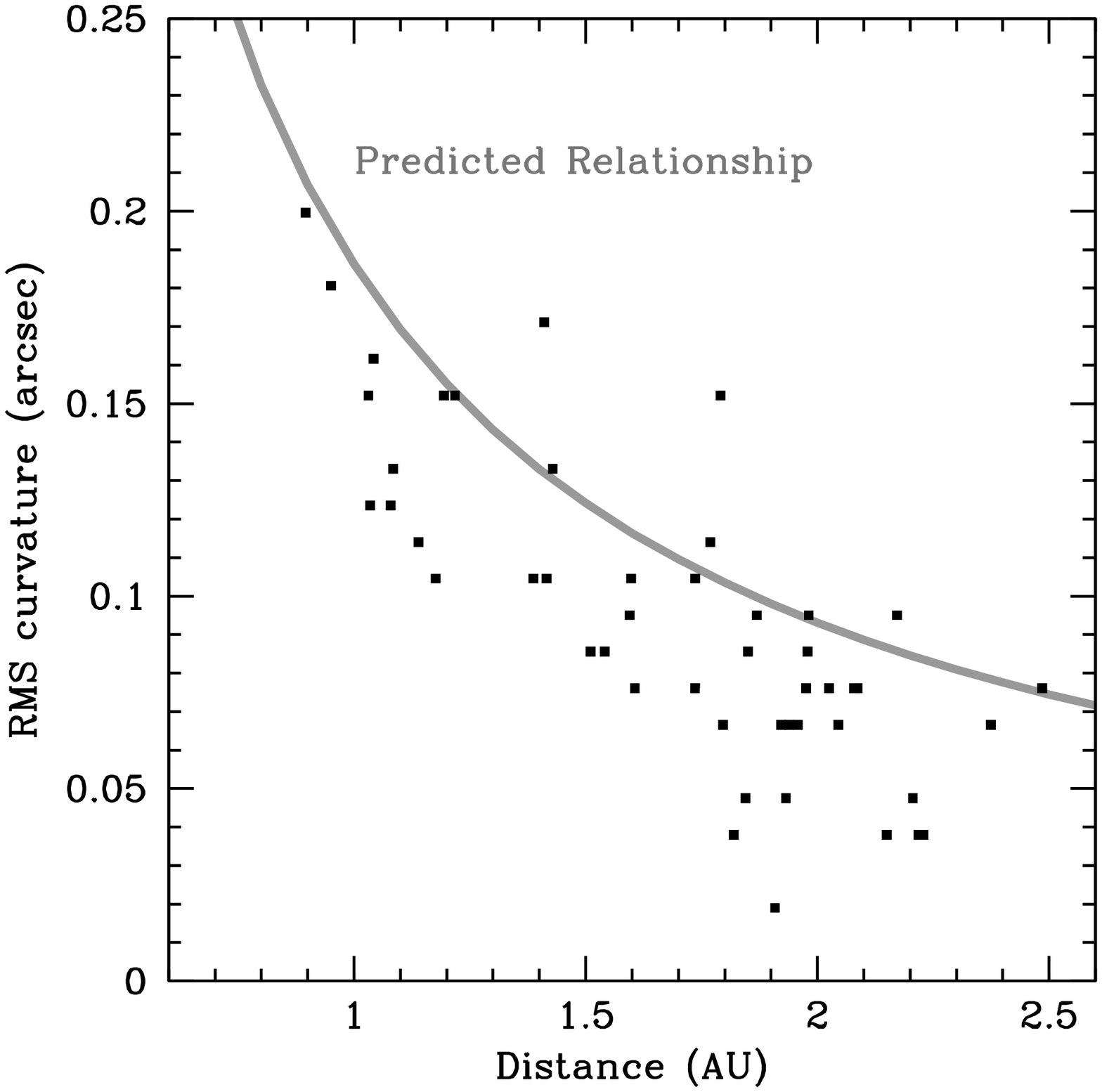}
\caption{\textbf{Left:} The curvature terms $\xi (t),\zeta (t)$ from Equation \ref{eq:curve2} for
an asteroid at a distance of 2 AU in our April 19 data. The projected rotational velocity of the
observer peaks at zero hour angle (near the center of the observing sequence); hence the asteroid's
retrograde (westward) angular velocity peaks at that time. \textbf{Right:} Measured rms curvature amplitudes 
(that is, the root mean of $\xi^2 (t) + \zeta^2 (t)$) for asteroids brighter than $R=20.5$ in our April 19 data,
plotted against distance and compared to the predicted (not fit) relationship from Equations \ref{eq:curve1} 
and \ref{eq:curve2}. Despite large scatter and a small systematic offset, the predicted $1/d$
dependence of curvature amplitude on distance is clearly seen, and the RMS error from this relationship is
only 0.03 arcsec. In principle, curvature measurements with
digital tracking data enable a measurement of an asteroid's geocentric distance based on only one night's data.
\label{fig:curve}}
\end{figure}

Averaged over the span of a digital tracking integration, $\Delta \alpha(t)$ and $\Delta \delta(t)$ contribute a constant offset to the measured position of an asteroid, and their first time derivatives $\dot {\Delta \alpha}(t)$ and $\dot {\Delta \delta}(t)$ produce a constant offset (i.e. the RRV signal itself) to its measured angular velocity. Neither of these constant offsets concerns us here: we wish to isolate and remove only the component that represents curvature and acceleration.  Thus, we create new functions $\xi(t)$ and $\zeta(t)$ that behave exactly like $\Delta \alpha (t)$ and $\Delta \delta (t)$ except that over the span of our digital tracking integrations, their means and the means of their first time derivatives are zero:

\begin{equation} \label{eq:curve2}
\begin{array}{lcl}
\xi (t) &=& \Delta \alpha (t) - \langle \dot {\Delta \alpha} (t)\rangle (t - \langle t\rangle) - \langle \Delta \alpha (t) \rangle \\
\\
\zeta (t) &=& \Delta \delta (t) - \langle \dot {\Delta \delta} (t) \rangle (t - \langle t \rangle ) - \langle \Delta \delta (t) \rangle
\end{array}
\end{equation}

\noindent where the notation $\langle \rangle$ denotes the time-average of the enclosed quantity over the span of a digital tracking integration. The left panel of Figure \ref{fig:curve} presents Equation \ref{eq:curve2} in graphical form, plotting the curved track of an asteroid in $\xi,\zeta$ space over the course of our April 19 observations. The calculation of $\xi$ and $\zeta$ of course requires knowledge of the asteroid's distance $d$, but the 2\% accurate values from Section \ref{sec:initdist} are more than sufficient for this purpose.

Thus, we adjust the methodology described in Section \ref{sec:angdigi} for measuring precise angular velocities from digital tracking data: we apply additional shifts equal to $-\xi (t), -\zeta(t)$ to each postage stamp image before the final stack. This straightens the asteroid's track so that curvature can no longer influence the calculated angular velocity. The distance can be recalculated based on the resulting new, de-biased angular velocities. Note that because Equation \ref{eq:curve2} subtracts \textit{average} reflex velocities, we must now use averaged rather than linear-fit values for $v_{rot}$ (Section \ref{sec:angdigi} and Figure \ref{fig:proj1} illustrate how the velocities differ). 

Compared to the calculation in Section \ref{sec:initdist}, our curvature correction reduces the mean absolute error of the distances for known objects from 0.042 AU to 0.032 AU, and the mean absolute fractional error from 2.2\% to 1.6\%. More significantly, the new calculation changes the weighted average signed fractional error from $-0.0179 \pm 0.0017$ to $-0.0028 \pm 0.0017$. The systematic offset is thus reduced by a factor of six, and the new value of -0.3\% is statistically consistent with zero. 

\begin{figure} 
\plottwo{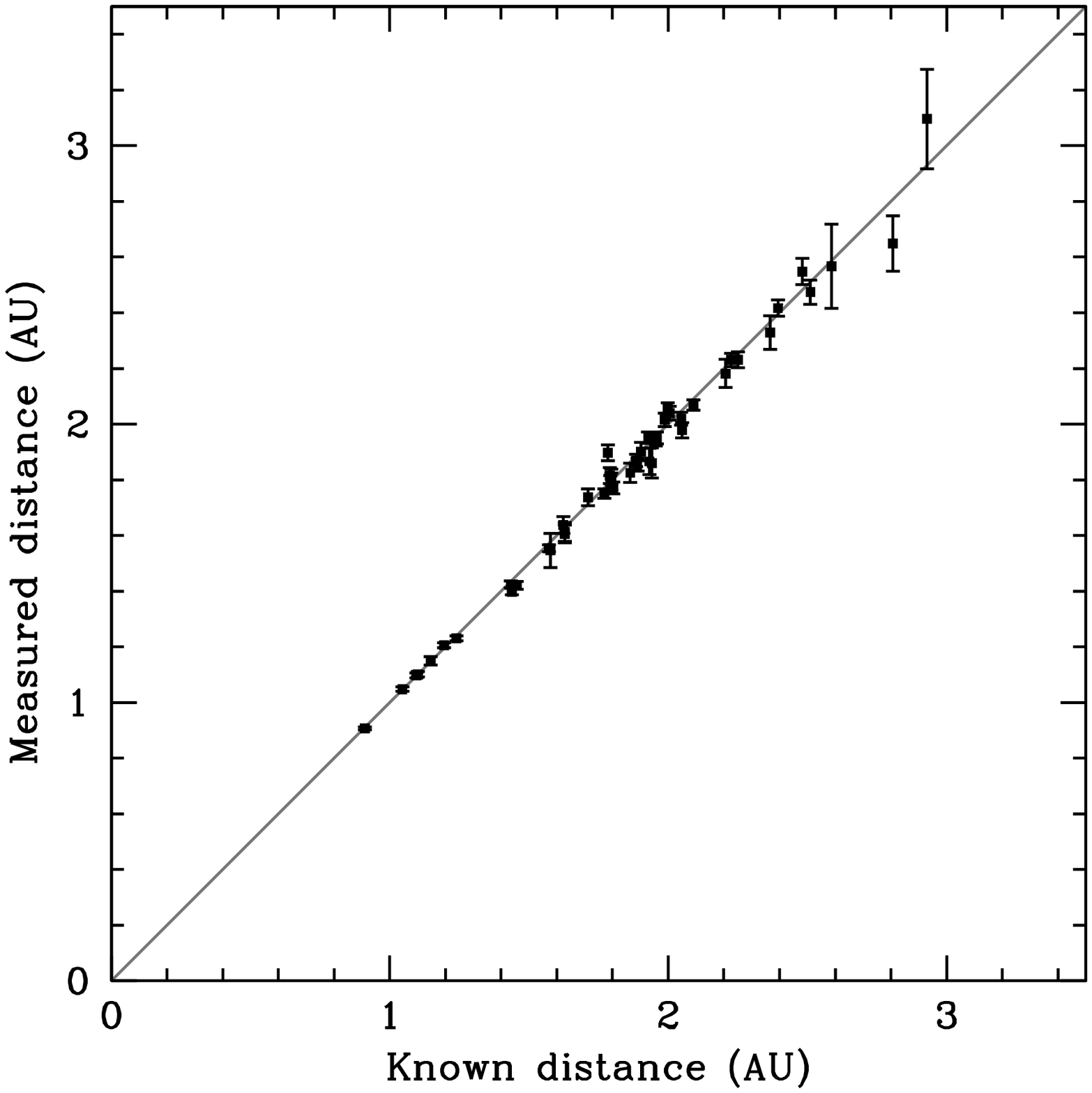}{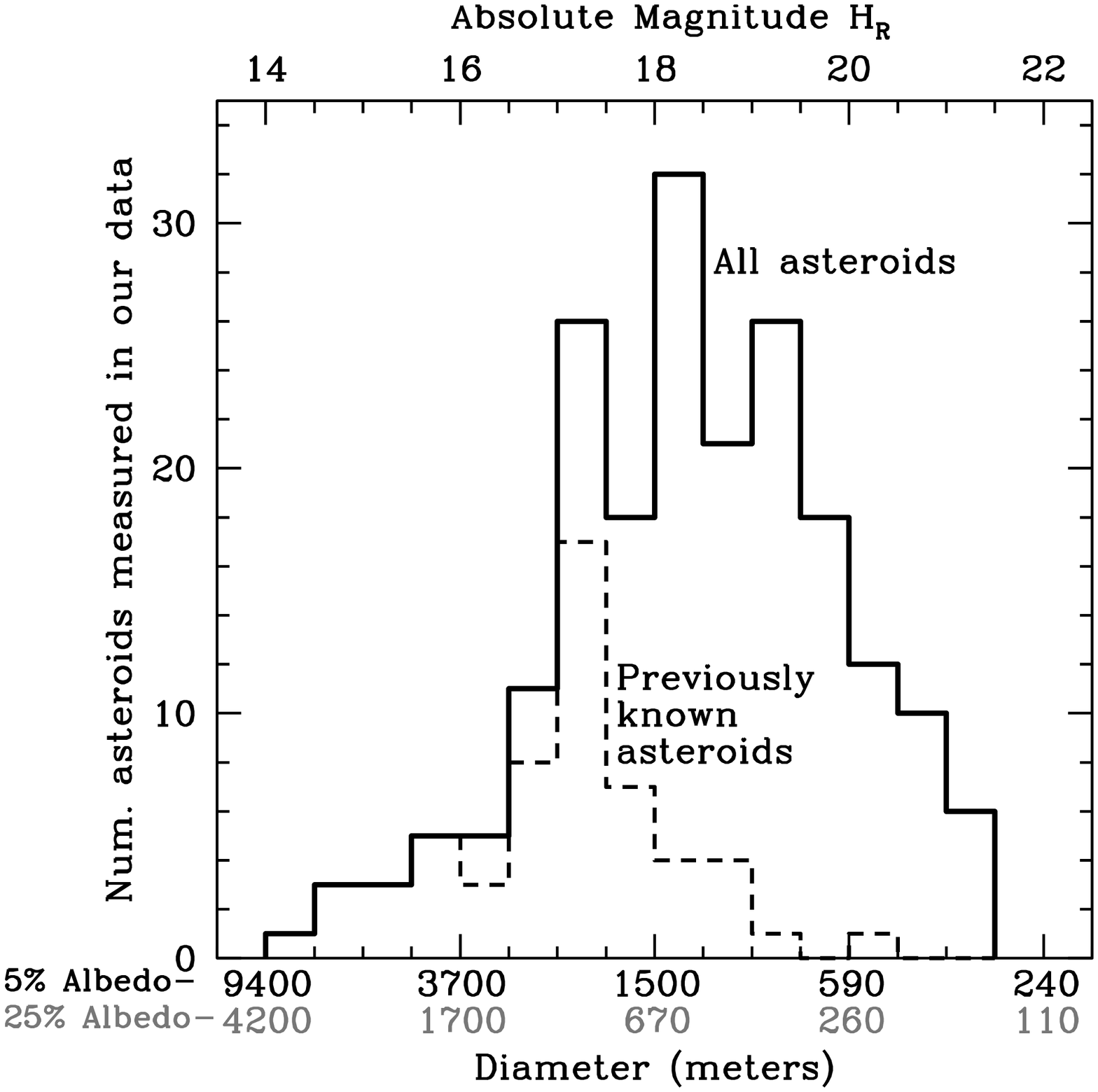}
\caption{\textbf{Left:} The precision of our final RRV distance measurements for known objects. The line is the 1:1 correspondence and is not a fit: our distances are calculated from basic geometry and known quantities describing the Earth, not based on a fit to known objects in our data. As expected given the $1/d$ dependence of the RRV signal, our measurements are most precise for the nearest objects. \textbf{Right:} Histograms of the apparent magnitudes (and corresponding physical diameters) for all asteroids (solid line) and for previously known objects (dashed line) measured in our data. The current census of the main belt becomes substantially incomplete at a diameter of about 2 km. By contrast, we have detected dozens of new asteroids in the 200-500 meter size range.
\label{fig:distmag}}
\end{figure}

The left panel of Figure \ref{fig:distmag} illustrates the precision of our measured distances for known objects. We emphasize that these distance measurements were entirely calculated from first principles and known quantities describing the Earth. We have used the known asteroids in our data to quantify our errors, but not to determine any of the parameters used in the distance calculation. 

\subsection{Measurements of Track Curvature within a Single Night} \label{sec:one_approx}

Beyond removing systematic errors in our measurements of asteroids' mean angular velocities, we can in principle use Equation \ref{eq:curve2} to \textit{measure} the curvatures of asteroid tracks within just a single night's data. This is an extremely challenging measurement because the curvatures are so small: the RMS curvature amplitude, defined as the root mean of $\xi^2 (t) + \zeta^2 (t)$, is less than 0.1 arcsec for an asteroid at 2 AU. We attempt to measure it using our curvature-corrected postage-stamp analysis as described in Section \ref{sec:two_precise}, but instead of calculating  $\xi (t)$ and $\zeta (t)$ based on a previously-calculated approximate distance, we probe a range of RMS curvature amplitudes for each asteroid.  For each trial curvature amplitude we calculate $\xi (t)$ and $\zeta (t)$, apply the appropriate shifts to the postage stamp images, and perform a quadratic fit to flux as a function of angular velocity as described in Section \ref{sec:angdigi}. We compare the peak asteroid flux values produced by the quadratic fits for the different trial curvature amplitudes, and identify the curvature amplitude that yields the highest peak asteroid flux. This should be the curvature value that resulted in all of the individual asteroid images being most accurately registered. As the curvature amplitudes map uniquely to distances (Equations \ref{eq:curve1} and \ref{eq:curve2}), our identification of the best-fit amplitude constitutes a distance measurement, in principle. However, the uncertainty of curvature amplitudes measured from the current data set is large enough that we do not choose to calculate individual curvature-based distances.

Instead, we plot the best-fit curvature amplitudes as a function of the known RRV distances calculated in Section \ref{sec:two_precise}. A clear inverse relationship emerges, as predicted by Equation \ref{eq:curve1}. The relationship gets tighter when we reject the minority of asteroids with out-of-range best-fit curvature values and all asteroids fainter than $R=20.5$ mag, which cannot be measured as accurately. The 48 objects meeting these criteria in our April 19 observations are plotted in the right-hand panel of Figure \ref{fig:curve}, which shows the RMS curvature amplitude as a function of geocentric distance in AU, and compares the data with the predicted curve from Equations \ref{eq:curve1} and \ref{eq:curve2}.  Although there is a mild systematic offset, the slope of the best linear fit to curvature as a function of $d^{-1}$ for these 48 asteroids is 9$\sigma$ significant, matches the predicted value to within 0.4$\sigma$, and has an intercept only 1.4$\sigma$ away from the predicted value of zero. The RMS error of the measured curvature amplitudes relative to their predicted values is only 0.03 arcsec. Thus we have effectively performed astrometry of moving, 20th magnitude objects with a precision of 30 milli-arcsec using only an 0.9m telescope.

In principle, our success at measuring asteroid curvatures within a single night's data means that digital tracking can be used to measure the geocentric distances to asteroids based on observations from only one night. However, our current curvature measurements have uncertainties that are too large to yield useful distances for individual objects: we have been able to demonstrate the efficacy of our curvature-measuring methodology only by recourse to an ensemble of 48 asteroids with distances known by other means. With a larger telescope, more accurate curvature measurements of fainter asteroids would be possible, especially in good seeing. These could conceivably yield meaningful distance measurements for unknown main-belt asteroids over just a single night, albeit only for objects considerably brighter than the digital tracking detection limit. By contrast, the two-night RRV calculations of Sections \ref{sec:initdist} -- \ref{sec:two_precise} yield precise distances even for faint objects just above the detection threshold. There is therefore little reason to use the curvature method for asteroids measured on more than one night.

%Where two nights of data are available, immensely more accurate distances may be calculated by the methods of Sections \ref{sec:initdist} -- \ref{sec:two_precise} without directly measuring curvature --- even for faint objects just above the detection threshold.

However, where only one night's data is available, distances based on digital-tracking curvature measurements could be very valuable. This is particularly true in at least two cases that may arise in future surveys.  The first is the case of an NEO survey where some objects --- perhaps due to their fast motion --- have inadvertently been measured on only one night. Being much closer to the Earth than MBAs, NEOs will show much larger and more easily measurable curvature, which is likely to yield useful distances. Such distances could enable the inclusion in a statistical analysis of objects that were otherwise unusable due to their one-night status. The second case is the one-night detection of a nearby NEO moving at nearly the same space velocity as Earth, such that its slow angular velocity mimics that of a much more distant object. This scenario is statistically rare but troubling, because such an object would likely be overlooked in an NEO survey and yet could be an incoming Earth-impactor. If the discovery survey used digital tracking, a curvature analysis would reveal the object's actual, very small geocentric distance. We note in closing that curvature measurements can in principle be performed on asteroids detected with conventional methods rather than digital tracking, but they will normally be less accurate because of the smaller number of images available.

\subsection{Distances and Sizes of Detected Asteroids} \label{sec:distsize}

While our current data only suggest the possibility of curvature-based distance measurements from a single night, they allow precise two-night distance measurements from rotational reflex velocity (Section \ref{sec:two_precise}). These precise distance measurements enable us to calculate the absolute magnitudes of each of our detected asteroids, and hence their physical diameters modulo the uncertainty in albedo.  The histograms of these values are shown in the right-hand panel of Figure \ref{fig:distmag}. At least half of the asteroids we have measured are smaller than 1 km. By contrast, only a small fraction of known MBAs are in this size range. The smallest objects we have detected are 300 meters in diameter under the assumption of low, 5\% albedo, but could be as small as 150 meters if their albedo is 25\%. The statistics of MBAs in this size range are already known \citep{Gladman2009} based on a sample several times larger than we present herein. Given this, the detailed completeness analysis that would be required to convert our detections into a measurement of the SFD is not worthwhile. The aims of the current work and our companion paper \citet{digitracks} are instead the validation of the RRV technique and of digital tracking for asteroids, respectively. Similar observations using a larger telescope, however, would extend to much smaller objects and accurately measure the SFD of a previously unexplored size regime in the main belt.

\section{Conclusion} \label{sec:conc}

We have described how the reflex angular velocity of asteroids due to Earth's rotation can be used to determine the distances to main belt asteroids based on only two nights of observations. We refer to this as the rotational reflex velocity (RRV) method for measuring asteroid distances (Equation \ref{eq:newav01}). The required approximations are accurate to about $10^{-3}$ (Equation \ref{eq:err02}), and measurement uncertainties are typically 1-3\%. Such distances can be used to calculate precise size statistics of small main belt asteroids using a much smaller investment of time on a large telescope than has previously been required.  Accurate RRV distances may be calculated either from conventional asteroid-search observations, from data analyzed by the technique of digital tracking \citep{Zhai2014,digitracks}, or from any other specialized asteroid-observation technique (e.g. Milani et al. 1996 or Gural et al. 2005) that accurately measures the celestial coordinates of the asteroids.

We have tested RRV distance determination with a data set acquired on April 19 and 20, 2013 using the WIYN 0.9-meter telescope at Kitt Peak. Using measurements based on digital tracking, we have calculated distances to 197 asteroids in this data set. While the majority of these were new discoveries in our data, 48 of them are previously known objects with accurate orbits.  These allowed us to test the accuracy of rotational reflex distances. A preliminary analysis yielded distances with a mean fractional error of only 2.2\%. Without digital tracking, our detections would be confined to substantially brighter asteroids (or, the observations would require a larger telescope), but the accuracy of RRV distances for detected objects would likely be at least this good (see Section \ref{sec:measprec}).

We have identified a 1.8\% systematic error in our preliminary RRV distances. We have linked this error to the curvature of the asteroids' observed tracks that results from Earth's rotation. This curvature, combined with the non-uniform sensitivity of images obtained under conditions of different seeing and airmass, causes a slight bias in the angular velocity measurements obtained from our digital tracking image-stacks. Finding our initial distance measurements more than sufficient to predict curvature amplitudes, we have calculated the form and amplitude of the curvature for each of our asteroids. We have applied a curvature correction to our image stacks to linearize the asteroid motions. Re-calculating the distances using angular velocities from curvature-corrected image stacks reduces the systematic bias from 1.8\% to $0.3 \pm 0.2$\%, and thus eliminates it as a significant effect.  The corrected distances have a mean fractional error of only 1.6\% for the 48 known objects in our data. Note that such a curvature correction would not be needed for RRV distances based on asteroids detected on discrete images. The calculation of accurate RRV distances would therefore be simpler for conventional asteroid detections, although digital tracking can detect much fainter objects.

Besides correcting our digital tracking angular velocity measurements for the curvature due to Earth's rotation, we have attempted to measure this curvature by creating several different image stacks for each asteroid at a range of different curvature amplitudes.  As the curvature amplitude is proportional to $1/d$, this constitutes a rotational reflex velocity measurement of an asteroid's distance based on just a single night's data.  We find that we can indeed measure the curvature amplitudes for asteroids brighter than $R = 20.5$ mag with a precision of about 30 milli-arcsec. The measured curvature values for 48 asteroids brighter than this limit show the expected $1/d$ dependence at 9$\sigma$ significance, and the best-fit slope is within 0.4$\sigma$ of the predicted value. While distance measurements based on single-night curvature amplitudes are too noisy to be useful for asteroids in our current data set (and are not needed given our highly precise two-night values), single-night distances could be valuable in specific cases for future surveys. In particular, curvature measurements can identify nearby objects moving at slow angular velocities characteristic of a much more distant population.

Our precise distances from two-night RRV measurements allow us to calculate absolute magnitudes and hence approximate diameters for our newly discovered asteroids. Our faintest objects have $H_R \sim 21.5$ mag and hence diameters of 130--300 meters depending on their albedo. While the current census of the main belt becomes substantially incomplete at a diameter of about 2 km, we have detected dozens of new asteroids in the 200--500 meter size range with an 0.9m telescope.

\section{Acknowledgments} 
Based on observations at Kitt Peak National Observatory, National Optical Astronomy Observatory (NOAO Prop. ID: 2013A-0501; PI: Aren Heinze), which is operated by the Association of Universities for Research in Astronomy (AURA) under a cooperative agreement with the National Science Foundation.

This publication makes use of the SIMBAD online database,
operated at CDS, Strasbourg, France, and the VizieR online database (see \citet{vizier}).
We have also made extensive use of information and code from \citet{nrc}. 
We have used digitized images from the Palomar Sky Survey 
(available from \url{http://stdatu.stsci.edu/cgi-bin/dss\_form}),
 which were produced at the Space 
Telescope Science Institute under U.S. Government grant NAG W-2166. 
The images of these surveys are based on photographic data obtained 
using the Oschin Schmidt Telescope on Palomar Mountain and the UK Schmidt Telescope.

Facilities: \facility{0.9m WIYN}

\end{document}